 \documentclass[useAMS,usenatbib]{mn2e}
 \usepackage{amssymb}
 \usepackage{graphicx}
 \usepackage{color}
 \usepackage{amsmath}
 
 \usepackage[T1]{fontenc}                                                                                                        
\usepackage{ae,aecompl}

 \usepackage{enumerate}
 \usepackage{hyperref}
 
\def\RLC{R_{\rm LC}}

\def\beq{\begin{equation}}
\def\eeq{\end{equation}}
\def\Eq{Equation}
\def\Eqs{Equations}
\def\Sect{Section}

\def\epB{\epsilon_B}
\def\epth{\epsilon_{\rm th}}

\def\Prad{P_{\rm rad}}
\def\Pgas{P_{\rm gas}}

\def\tvisc{t_{\rm visc}}
\def\vK{v_{\rm K}}

\def\Rph{R_\star}
\def\Rd{R_d}
\def\dM{\dot{M}}
   \def\Uth{U}
   \def\Eout{{\cal E}}
   \def\Mout{{\cal M}}
 \def\Jout{{\cal J}}
   \def\tout{{\cal T}}
\def\geff{g}
\def\Fm{F_m}
\def\LE{L_{\rm E}}

\def\dEem{\dot{E}_{\rm P}}
\def\dEm{\dot{E}_{\rm mat}}
\def\ld{l_D}
\def\Rd{R_D}
\def\Lph{L_\star}
\def\tph{t_\star}

\def\effph{\eta_\star}
\def\kappaT{\kappa_{\rm T}}
\def\RA{R_{\rm A}}
\def\vA{v_{\rm A}}

\def\Fop{\Psi_{\rm op}}
\def\Rdec{R_{\rm dec}}

\newbox\grsign \setbox\grsign=\hbox{$>$} \newdimen\grdimen \grdimen=\ht\grsign
\newbox\simlessbox \newbox\simgreatbox \newbox\simpropbox
\setbox\simgreatbox=\hbox{\raise.5ex\hbox{$>$}\llap
     {\lower.5ex\hbox{$\sim$}}}\ht1=\grdimen\dp1=0pt
\setbox\simlessbox=\hbox{\raise.5ex\hbox{$<$}\llap
     {\lower.5ex\hbox{$\sim$}}}\ht2=\grdimen\dp2=0pt
\setbox\simpropbox=\hbox{\raise.5ex\hbox{$\propto$}\llap
     {\lower.5ex\hbox{$\sim$}}}\ht2=\grdimen\dp2=0pt
\def\simgt{\mathrel{\copy\simgreatbox}}
\def\simlt{\mathrel{\copy\simlessbox}}

\begin{document}

\title[Outbursts from white-dwarf mergers]
{Magnetically powered outbursts from white dwarf mergers}
\author[A. M. Beloborodov]
       {Andrei M. Beloborodov\thanks{E-mail: amb@phys.columbia.edu} \\
           Physics Department and Columbia Astrophysics Laboratory
 Columbia University, 538  West 120th Street New York, NY 10027  
                }

% \date{\today}
\date{Received / Accepted}

\maketitle

\label{firstpage}

\begin{abstract}
Merger of a white dwarf binary creates a differentially rotating object which
is expected to generate strong magnetic fields. 
Kinetic energy stored in differential rotation is partially dissipated 
in the magnetically dominated corona, which forms a hot variable outflow
with ejection velocity comparable to $10^9$~cm~s$^{-1}$.
The outflow should carry significant mass and energy 
for hours to days, creating an expanding fireball with the following features. 
(i) The fireball is initially opaque and its internal energy is dominated by the trapped 
thermal radiation. The stored heat is partially 
converted to kinetic energy of the flow (through adiabatic cooling) and partially 
radiated away. 
(ii) Internal shocks develop in the fireball and increase its radiative output.
(iii) A significant fraction of the emitted energy is in the optical band. As a result,
a bright optical transient with luminosity $L\sim 10^{41}-10^{42}$~erg~s$^{-1}$
and a characteristic peak duration comparable to 1~day may be expected from the 
merger. In contrast to classical novae or supernovae, the transient does not involve 
nuclear energy. The decay after its peak reflects the damping of differential rotation 
in the merger remnant. Such outbursts may be detected in the local Universe 
with current and upcoming optical surveys.
\end{abstract}

% \keywords{plasmas --- stars: magnetic fields, neutron }
\begin{keywords}
binaries: close ---
magnetic fields --- MHD ---  radiation mechanisms: general
--- stars: coronae, rotation, winds, outflows --- supernovae: general
--- white dwarfs 
\end{keywords}

\maketitle

%#############################################################

\section{Introduction}

Population synthesis models suggest the birth rate of binary white dwarfs (WD) 
in our Galaxy comparable to 0.05~yr$^{-1}$, and a large fraction of these
binaries, perhaps a half, are expected to merge in less than a Hubble time
(Nelemans et al. 2001).
The possibility of detecting gravitational waves 
could make the WD mergers particularly interesting sources for future observations.
About 30 tight WD binaries have already been found (Kilic et al. 2012);
however, no mergers events have yet been identified. It was suggested that
some of them may be associated with thermonuclear supernovae (SN Ia, 
  Iben \& Tutukov 1984; Webbink 1984).  The estimated rate of WD mergers
is comparable to that of SN~Ia
(Badenes \& Maoz 2012), supporting their candidacy for SN~Ia progenitors.
Many of the mergers are, however, unable to ignite a thermonuclear explosion
--- the possibility of ignition depends on
the mass ratio of the binary and its chemical composition (Dan et al. 2012). 

In this paper, we discuss WD mergers that do not explode or explode with 
a significant delay (Raskin et al. 2009). Using simple estimates,
we argue that even without the liberation of nuclear energy
the mergers should eject bright fireballs detectable in optical surveys.

\subsection{Post-merger object}

The merger forms an axisymmetric, rapidly rotating central core
surrounded by a less massive debris disc 
(e.g. Benz et al. 1990;  Mochkovitch \& Livio 1990;
Rasio \& Shapiro 1995; Guerrero et al. 2004; Yoon et al. 2007; 
Lor\'en-Aguilar et al. 2009; Raskin et al. 2012).
The energy budget of this nascent object is
\beq
\label{eq:E0}
   E_0=\frac{GM^2}{R}\sim 3\times 10^{50} 
     \left(\frac{M}{M_\odot}\right)\left(\frac{R}{10^9{\rm cm}}\right)^{-1} {\rm erg},
\eeq 
where $M\sim M_\odot$ and $R\sim 10^9$~cm are the characteristic mass and
radius of the merger remnant. 
The remnant has three important features:

(1) Fast rotation. The orbital angular momentum of the binary 
system is inherited by the post-merger object. Its characteristic angular velocity 
$\Omega$ is comparable to the maximum (break-up) angular velocity 
$\Omega_{\max}=(GM/R^3)^{1/2}$.
The outer parts of the core and the surrounding disc have significant 
differential rotation, i.e. $\Omega$ varies with cylindrical
radius by $\Delta\Omega\sim\Omega$.

(2) The deep interior of the merger core can remain degenerate,
however its upper layers and the surrounding debris disc are strongly 
heated by shocks (e.g. Raskin et al. 2012),
which lifts the electron degeneracy and creates significant thermal pressure.
A large contribution to the pressure is made by radiation, $\Prad=aT^4/3$,
where $T$ is the temperature and $a$ is the radiation constant. Convection is 
likely to develop in the upper layers of the core and the debris disc.

(3) The remnant is expected to be strongly magnetized. Even if the pre-merger 
magnetic fields are weak, the differential rotation and convection 
quickly generate strong fields. This process was recently studied by Ji et al. 
(2013) whose magnetohydrodynamic simulations show fields up to 
$10^{10}-10^{11}$~G.

The generated field is convenient 
to express using the ``magnetization parameter'' --- the ratio of magnetic energy 
$E_B\sim B^2 R^3$ (omitting a numerical factor $\sim 0.1$) to the total energy 
of the object $E_0\sim GM^2/R$,
\beq
\label{eq:epB}
    \epB\equiv \frac{B^2R^4}{GM^2}.
\eeq
In this paper we focus on the low-density corona formed around the merger, 
and hereafter $\epB$ corresponds to the magnetic field in the corona.

\subsection{Analogies with other astrophysical differential rotators}

One close analogy is provided by neutron-star (NS) mergers. 
They received significant attention as major sources of gravitational waves;
they are also thought to produce short gamma-ray bursts (GRBs, e.g. Piran 2004). 
A strongly magnetized corona and outflows are believed to develop in NS mergers.
Although they differ from WD mergers in many respects --- e.g. they have a 
smaller size (by a factor of $\sim 10^3$) and a much higher cooling rate
due to neutrino emission, --- basic physics of magnetic field amplification is similar. 
Both WD and NS mergers create an object that can be described as an  
excellent conductor with fast differential rotation and convection; the Rossby 
numbers for the two cases are comparable.
Similar dynamo processes are also expected in proto-neutron stars formed in 
stellar collapse. Differential rotation generates a strong toroidal magnetic field, 
which is buoyant and forms a magnetically dominated corona
(e.g. Ruderman et al. 2000; Spruit 2008). 
Unlike proto-neutron stars, the WD merger is not cooled by neutrino emission 
and does not become solid; the lifetime of its differential rotation and coronal
activity is controlled by the effective viscosity due to magnetic stresses and 
turbulent diffusion.

Rich observational data are available for another class of fast magnetized rotators 
---  pre-main-sequence stars (protostars). They have a similar, 
fast-rotating core surrounded by a Keplerian disk.  
These objects are significantly bigger
in size than WD mergers, by a factor of $\sim 10^2$. Their characteristic age is 
$\sim 10^5$~yr and their inferred magnetization is modest.
Nevertheless, the observed rotation-powered activity of protostars 
(e.g. Montmerle et al. 2000; Getman et al. 2008) can provide a 
useful analogy. They show a strong coronal activity --- X-ray flares
triggered by reconnection events. Their average X-ray luminosity is 
a factor $10^3-10^5$ larger than that of the sun (e.g. Feigelson et al. 2007).
More generally, 
accretion discs are observed 
in protostars, X-ray binaries, and quasars. These canonical differential 
rotators are known to produce magnetized jets and strong nonthermal 
emission.

Amplification of magnetic fields is also suspected in stellar mergers, 
one of which has been caught by recent observations and gave rise to the 
red nova V1309 Scorpii (Tylenda et al. 2011). It provides a possible mechanism 
for formation of magnetic Ap/Bp stars
(Soker \& Tylenda 2007; Ferrario et al. 2009; Tutukov \& Fedorova 2010).
The WD merger is intermediate between the stellar mergers and neutron-star 
mergers in terms of size, active lifetime and energy budget of the remnant.

%#################################################################

\section{Active corona}

\subsection{Formation}

The merger remnant maintains hydrostatic equilibrium on the sound crossing
timescale,
\beq
   t_0=\frac{R}{v_0}=\left(\frac{R^3}{GM}\right)^{1/2}
   =\Omega_{\max}^{-1}\sim 3 {\rm ~s}. 
\eeq
Its density profile is controlled by the distribution
of entropy generated by shocks in the merger and subsequent viscous
dissipation. 
The low-density upper layers 
are expected to have higher entropy per unit mass, and supported mainly by 
radiation pressure.\footnote{The average entropy per unit 
     mass in WD mergers corresponds to $\Prad\sim \Pgas$. Upper layers with 
     higher entropy have $\Prad\gg\Pgas$.}
The object is in differential rotation and will tend to 
redistribute its angular momentum on a viscous 
timescale  $\tvisc\simgt 10^4$~s (Shen et al. 2012; Ji et al. 2013).
This timescale depends on the value of viscosity created by magnetic 
fields (and turbulence) in the remnant; viscous stress may be parameterized
as $T_{r\phi}= \alpha P$ where $P\simlt GM^2/R^4$ is the characteristic 
pressure. The timescale $\tvisc\sim 10^4$~s corresponds to 
$\alpha\sim 0.01$.
In Section~3  below we mainly focus on early times $t<\tvisc$, 
when differential rotation is still strong.
At this stage, magnetic fields are amplified and buoyantly emerge from 
the remnant on a timescale that is intermediate between $t_0$ and $\tvisc$.
An active magnetically dominated corona must be sustained 
around the differential rotator.

Strong magnetic fields should be generated in the merger debris disc
as well as in the upper layers of the central core. The disc
has the initial scale-height
$H/r\sim 0.1$ (Raskin et al. 2012). Its heating and further evolution develops on 
the viscous timescale $\tvisc \sim \alpha^{-1}(H/r)^{-2} \Omega^{-1}(r)\sim 10^4$~s.
The resulting super-Eddington accretion disc becomes thick and prone to 
outflow formation.
The net energy released on the viscous timescale is comparable to the kinetic 
energy of the rotating matter. This matter, which was initially gravitationally
bound, will remain bound if the generated heat is distributed strictly in proportion 
to mass density, giving a sound speed $c_s\sim \vK=(GM/r)^{1/2}$. 
Then the disc can 
evolve into a bound, radiation pressure-supported, quasi-spherical structure, 
growing in size and slowing down its rotation (Shen et al. 2012). 
However, the uniform heating is unlikely. Magnetic fields generated by the 
magneto-rotational instability are buoyant and expected to deliver and
dissipate energy in the upper layers of lower density. As a result,
a fraction $f$ of the disc mass is heated to $c_s\simgt \vK$ and
becomes gravitationally unbound. For instance, $f\sim 10^{-2}$ will give  
an outflow of total mass $\sim 10^{-3}M_\odot$ and energy $\sim 10^{48}$~erg. 

Magnetic activity of the post-merger object is 
demonstrated by recent numerical simulations of Ji et al. (2013).
They find that the magnetic energy of the remnant at its peak exceeds 
$10^{48}$~erg (which corresponds to a space-average $B\sim 10^{11}$~G) 
and a significant mass $\Mout\sim 10^{-3}M_\odot$ is ejected from 
the system over the run time of their simulations, $t=2\times 10^4$~s.

Hereafter ``corona'' refers to the magnetically dominated
region of radius $r\sim R$ coupled to the central core or the disc.
We will assume that the magnetization parameter of the corona,
as defined in \Eq~(\ref{eq:epB}), satisfies $\epB>10^{-5}$ which approximately
corresponds to the magnetic field $B> 2\times 10^{9}$~G.

Unlike the solar corona, the coronae of WD mergers are opaque to radiation. 
The photosphere lies at a large radius in the outflow 
zone which will be described below.

%#################################################################

\subsection{Dissipated energy}

Shearing of the footprints of coronal magnetic field lines 
by differential rotation and convective motions in the remnant
repeatedly twists the field lines to the threshold 
of instability, leading to magnetic flares and field-line opening similar to 
coronal mass ejection in solar flares. 
An upper limit for the dissipation rate in the corona 
is given by  $E_B\Omega$, where 
$E_B\sim 0.1 \epB GM^2/R$ is the magnetic energy of the corona, 
\beq
  L_c<L_{\max}\sim 0.1\epB \frac{GM^2}{R}\Omega \sim 10^{49} \epB {\rm ~erg~s}^{-1}.
\eeq
The corona may be approximately described as a force-free magnetic
configuration with dissipation localized in current sheets. 
Numerical simulations of dissipation are challenging, with results 
depending on the artificial (numerical) resistivity.
Simulations by Ji et al. (2013) suggest strong coronal dissipation during 
$\tvisc\sim 10^4$~s, with the average $L_c\sim 10^{44}$~erg~s$^{-1}$.

The magnetic energy around the remnant is continually pumped by the shearing 
motions of the field-line footprints frozen in the differential rotator, and a
quasi-steady balance between pumping and dissipation (flares) is maintained. 
As a result, a fraction of the rotational energy stored in the merger, 
$E_0\sim GM^2/R$, is gradually lost through the coronal dissipation and
outflow. Assuming that the most active phase of this process ends together
with strong differential rotation at $t\sim\tvisc$, the lost energy may be written 
as
\beq
   \Eout\sim L_c\tvisc.
\eeq
For numerical estimates we will adopt $\Eout \sim 10^{48}$~erg, similar 
to the results of Ji et al. (2013); note that it is less than one per cent 
of the merger energy $E_0$ given by \Eq~(\ref{eq:E0}).

The outflow energy is carried by matter and magnetic fields; the typical 
ejection velocity $v$ is a few times the escape velocity $v_0$ (see Section~2.4 
below), and the characteristic ejected mass is 
\beq
   \Mout\sim 10^{-3}M_\odot\, \Eout_{48} v_9^{-2}.
\eeq

\subsection{Density and temperature}

The ejection of mass $\Mout\sim 10^{-3}M_\odot$ during $\tvisc\sim 10^4$~s 
corresponds to the average mass loss rate $\dM\sim 2\times 10^{26}$~g~s$^{-1}$.
Let $t_c$ be the residence time of gas in the corona before it is ejected;
$t_c$ cannot be shorter than the sound crossing time $t_0=R/v_0\sim 3$~s
or the timescale for field-line twisting $\Delta\Omega^{-1}\simgt t_0$.
The characteristic mass density of the corona is related to $t_c$ by
\beq
   \rho\sim \frac{\dM t_c}{4\pi R^3}
         \sim 1\, \dM_{26}\,\left(\frac{t_c}{100 \rm ~s}\right) {\rm ~g~cm}^{-3}.
\eeq
It is much smaller than the interior density of the merger remnant 
$\rho\sim M/R^3\sim 10^6$~g~cm$^{-3}$.

The hydrostatic scale-height of the corona $H$ is comparable to the 
remnant radius $R\sim 10^9$~cm. 
The plasma confined in the magnetic field is free to expand 
along the field lines, and the condition $H\sim R$ corresponds 
to the sound speed $c_s\sim (P/\rho)^{1/2}$ being comparable to the virial 
velocity,
\beq
\label{eq:cs}
   c_s\sim v_0=\left(\frac{GM}{R}\right)^{1/2}\sim 4\times 10^8 {\rm ~cm~s}^{-1}.
\eeq
This implies a relation between the characteristic mass density $\rho$ and 
pressure $P=U/3$ in the corona: $P\sim \rho v_0^2$.
The pressure is dominated by radiation, $P=aT^4/3$, which gives
\beq
   T\approx \left(\frac{3\rho v_0^2}{a}\right)^{1/4}\sim 10^8 \rho^{1/4} {\rm ~K},
\eeq
where $\rho$ is in units of g~cm$^{-3}$.

Using these estimates one can verify that the corona is 
completely opaque to radiation, and its thermal energy density and pressure
are dominated by blackbody photons.
Gas pressure $\Pgas\approx\rho kT/m_p$ (where $m_p$ is proton mass) 
contributes a small fraction to the total pressure $P$,
\beq
   \frac{\Pgas}{P}=\frac{\rho kT}{m_pP}\sim \frac{kT}{m_pv_0^2}
    \sim 5\times 10^{-2} \left(\frac{T}{10^8\rm ~K}\right).
\eeq

The thermal energy density of the corona $U=3P$ is supplied by dissipation 
of magnetic energy, and may be written as
\beq
   \Uth=\epth \frac{B^2}{8\pi}\sim\epth \epB\,\frac{GM^2}{R^4},
\eeq
where $\epth<1$. In a local dissipative region (a flare generated by 
an unstable current sheet), $\epth$ is not much below unity; the average
value of $\epth$ is much smaller. 
Note that $\rho \sim 1$~g~cm$^{-3}$ and $U\sim \rho v_0^2$ 
roughly corresponds to $\epth\epB\sim 10^{-6}$.

\subsection{Ejection velocity}

The outflow is gravitationally unbound and its minimum expected velocity is 
comparable to the virial velocity $v_0$. The outflow can be additionally
accelerated by the magnetic Lorentz force (equivalent to centrifugal 
acceleration if viewed in the frame co-rotating with the open magnetic field lines),
as described by the standard theory of magnetized winds 
(e.g. Lamers \& Cassinelli 1999). The acceleration is significant if the magnetic 
flux penetrating the outflow $\Fop$ (open flux) is sufficiently strong to enforce 
the flow co-rotation with angular velocity $\Omega$ to a large radius $\RA>R$.
Then the flow can be ejected with the velocity $v\sim \vA\sim\Omega\RA$.
Here $\RA$ is the Alfv\'en radius at which $B^2\sim 4\pi\rho v^2$; this
condition can be rewritten using the open magnetic flux $\Fop\sim Br^2$ and
mass flux $\dM\sim \rho v r^2$ at $r\sim\RA$,
\beq
   \frac{\Fop^2}{\RA^4}\sim 4\pi\frac{\dM}{\RA^2}\,\vA.
\eeq
Substitution of $\RA\sim\vA/\Omega$ gives the equation for $\vA$, and one finds
that the outflow is ejected with 
\beq
   v\sim \left(\frac{\Omega^2\Fop^2}{4\pi \dM}\right)^{1/3}
     \approx 10^9 \,\Omega_{-1}^{2/3}\, \Psi_{\rm op,28}^{2/3}\,\dot{M}_{26}^{-1/3}
       {\rm ~cm~s}^{-1}.
\eeq
This standard estimate assumes $v_0<v<c$.
If $v$ approaches $c$, the relativistic wind theory should be used
(Michel 1969), however this does not occur in our fiducial model with 
$\dM\sim 10^{26}$~g~s$^{-1}$. The active corona can only open a 
fraction of the total magnetic flux of the post-merger object,
$\Fop<\Psi\sim BR^2\sim 10^{28}B_{10}R_9^2$~G~cm$^2$.
The condition $v>v_0$ requires $\Fop>10^{27}\dM_{26}^{1/2}$~G~cm$^2$.
In the numerical estimates below we adopt 
the outflow speed $v\sim 10^9$~cm~s$^{-1}$, 
a few times larger than $v_0$. Ji et al. (2013) find a comparable $v\sim 2v_0$.

\subsection{Spindown effect of the outflow}

The model outlined above assumes that the energy stored in differential 
rotation is mostly dissipated inside the remnant, and roughly 
one  per cent or less is dissipated in the corona and feeds the outflow. 
The outflow also carries away angular momentum, 
\beq
  \Jout \sim \Mout \RA v\sim \frac{\Mout v^2}{\Omega},
\eeq
which is a small fraction of the total angular momentum
of the object, $J\sim MR^2\Omega$,
\beq
   \frac{\Jout}{J}\sim \frac{\Mout}{M}\,\left(\frac{v}{v_0}\right)^2
   \left(\frac{\Omega}{\Omega_{\max}}\right)^{-2}\ll 1.
\eeq 
This ratio is $\sim 10^{-2}$ for the typical parameters of our model.

The remnant can slowly lose its angular momentum to a weaker wind at much
longer times $t\gg\tvisc$, when the mass loss rate is reduced.
Note that even with conserved angular momentum the angular 
velocity of the remnant can decrease on the timescale $\tvisc$
as it viscously spreads into a bigger object (Shen et al. 2012).

%#################################################################

\section{Radiation from the fireball}

Consider a fireball of mass $\Mout\sim 10^{-3}M_\odot\approx 2\times 10^{30}$~g 
ejected during time $\tout\sim 10^4-10^5$~s with velocity $v\sim 10^9$~cm~s$^{-1}$. 
Its heat is comparable to its kinetic energy, 
$\Eout\sim \Mout v^2/2\sim 10^{48}$~erg. 
The heat can be lost through radiative diffusion and adiabatic cooling; on 
the other hand, it can be generated by delayed internal dissipation, in particular by 
internal shocks in the fireball. These processes and the resulting emission
are discussed below.

\subsection{Diffusion radius and photospheric radius}

Radiation tends to diffuse out of the fireball. In time $t$, radiation diffuses 
through the characteristic length defined by $\ld^2=Dt$, where $D=(3\kappa\rho)^{-1}c$ 
and $\kappa$ is the opacity. As a reference value for $\kappa$ one can use Thomson 
opacity $\kappaT\approx 0.2$~cm$^2$~g$^{-1}$; the actual $\kappa$ will be 
discussed below. The fireball density at radius $r=vt$ may be written as
\beq
  \rho=\frac{\dM}{4\pi r^2 v}\approx \frac{\Mout}{4\pi r^2 v \tout},
\eeq
where we assumed that the fireball is quasi-spherical (it is straightforward to 
extend the model to beamed outflows). This gives the diffusion length at radius $r$,
\beq
   \ld=\left(\frac{4\pi r^3 c}{3\kappa \dM}\right)^{1/2}.
\eeq
When $\ld$ approaches the fireball thickness $l_f=\min\{v\tout,r\}$ radiation 
is no longer trapped. The condition $\ld= l_f$ defines the characteristic 
``diffusion radius'' $\Rd$ where radiation escapes the fireball.
This radius is given by
\begin{eqnarray}
\label{eq:Rd}
  \Rd=\left\{ \begin{array}{ll}
      \displaystyle{ \left(\frac{3\kappa \tout\Mout v^2}{4\pi c}\right)^{1/3}
      = v\tout_1 \left(\frac{\tout}{\tout_1}\right)^{1/3} }, & \tout<\tout_1 \\
      \displaystyle{ \frac{3\kappa \dM}{4\pi c}   
      = v\tout_1 \left(\frac{\tout}{\tout_1}\right)^{-1} },   & \tout>\tout_1 
                  \end{array} 
          \right.
\end{eqnarray}
$\Rd$ is maximum if $\tout=\tout_1$,
\beq
   \tout_1=\left(\frac{3\kappa \Mout}{4\pi c v}\right)^{1/2} 
   \approx 6\times 10^4\, \left(\frac{\kappa}{\kappaT}\right)^{1/2}
   \Mout_{-3}^{1/2}v_9^{-1/2} {\rm ~s},
\eeq
where $\Mout_{-3}=\Mout/10^{-3} M_\odot$ and $v_9=v/10^9$cm~s$^{-1}$.
A typical diffusion radius is $\Rd\sim 3\times 10^{13}$~cm.

At a larger radius $\Rph$, the fireball becomes transparent to radiation,
\beq
\label{eq:Rph}
   \Rph\approx \left(\frac{\kappa\Mout}{4\pi}\right)^{1/2}
   \approx 2\times 10^{14}\, \left(\frac{\kappa}{\kappaT}\right)^{1/2}
   \Mout_{-3}^{1/2} {\rm ~cm},
\eeq
where we assumed $\tout<\Rph/v$. At late times $t>\Rph/v$ the photosphere
shrinks; its evolution is controlled by the decreasing $\dM$:
$\Rph=\kappa\dM/4\pi v$.

In the above numerical estimates for $\Rd$ and $\Rph$, we normalized the 
opacity $\kappa$ to its Thomson value 
$\kappaT\approx 0.2$~cm$^2$~g$^{-1}$.
The actual Rosseland mean opacity is comparable to $\kappaT$ at radii
$r\simlt\Rd$ and reduced at $r\sim 10^{14}$~cm (\Sect~3.4), which 
slightly reduces the photospheric radius.

\subsection{Effective temperature and optical luminosity}

For a given luminosity $L$ emitted by the fireball at a radius $r$, the effective
temperature of radiation is given by 
\beq
    T=\left(\frac{L}{4\pi r^2 \sigma}\right)^{1/4}
    \approx 1.9\times 10^4\, L_{42}^{1/4} r_{14}^{-1/2} {\rm ~K},
\eeq 
where $\sigma$ is the Stefan-Boltzmann constant. Assuming 
a quasi-thermal radiation spectrum near the photosphere, one can see that 
it will peak near the optical band if $L\sim 10^{41}-10^{42}$~erg~s$^{-1}$.
At smaller $r$ or higher $L$, one finds $kT>h\nu$ for the optical frequency 
$\nu\sim 6\times 10^{14}$~Hz. Then 
the luminosity in the optical band $L_O$ can be significantly smaller than
the total emitted luminosity $L$. Approximating the emission at $h\nu<kT$ by 
the Raleigh-Jeans formula, one finds
\beq
\label{eq:LO}
   L_O\sim 8\pi^2 r^2 \frac{\nu^3}{c^2}\,kT
        \approx 5\times 10^{41}\, L_{42}^{1/4} r_{14}^{3/2} {\rm ~erg~s}^{-1}.
\eeq

\subsection{Adiabatic cooling and internal shock heating}

In the absence of delayed internal dissipation, the expanding opaque fireball cools
adiabatically with adiabatic index $\gamma=4/3$. Its energy density is decreasing as 
$U\propto r^{-8/3}$, which corresponds to $T\propto r^{-2/3}$.
As the fireball expands from $r\sim R$ to the diffusion radius $\Rd$, its volume 
is increased as $(r/\Rd)^2$ and the total thermal energy is reduced as $(r/\Rd)^{-2/3}$.
Thus, the total energy emitted to distant observers is
\beq
   \Eout_{\rm em}\sim \left(\frac{\Rd}{R}\right)^{-2/3}\Eout\sim 10^{-3}\,\Eout.
\eeq
It is emitted on a timescale $t\sim\Rd/v$ if this time is longer than $\tout$,
which gives luminosity
\beq
   L\sim  \left(\frac{\Rd}{R}\right)^{-2/3} \frac{v\,\Eout}{\Rd}\sim 10^{40}-10^{41}{\rm ~erg~s}^{-1}.
\eeq
A more detailed estimate takes into account the additional cooling that occurs
when the fireball is magnetically accelerated to $v>v_0$; then there is an 
additional cooling factor $(v/v_0)^{-4/3}$.

The luminosity emitted by the fireball at earlier times, when its radius $r<\Rd$,
may be estimated as
\beq
    L\sim \frac{4\pi r^2 \ld U}{t},  \qquad r<\Rd.
\eeq
One then finds that the bolometric luminosity is slightly higher at earlier times, 
$L\propto t^{-1/6}$. Luminosity in the optical band is smaller at $t<\Rd/v$;
it reaches its peak at $t\sim \Rd/v$. The rise of the optical luminosity toward the
peak can be estimated using \Eq~(\ref{eq:LO}) and 
$L^{1/4}\propto t^{-1/24}\approx const$, which gives $L_O\propto t^{3/2}$.

These estimates assume passive adiabatic cooling of the expanding fireball.
In reality, it is likely to experience internal heating in a broad range of radii.
The outflow is created by the variable corona of the differential rotator,
and its ejection velocity $v$ can vary by $\Delta v\sim v$ on a broad range of 
timescales $\Delta t_{\rm var}$, 
from $R/v_0$ to the age of the remnant. As the outflow cools, its variable velocity 
profile leads to internal supersonic motions and formation of shocks at radius 
$r\sim v^2 \Delta t_{\rm var}/\Delta v$. Then part of 
the energy lost to adiabatic cooling is converted back to heat. 
The shock-heated plasma again adiabatically cools, and new
shocks can form. The shock heating can continue to large radii, even approaching
the photospheric radius $\Rph$. Note the possibility of a gradual increase
of the ejection velocity on the long timescale $\tout\sim\tvisc$, as at late times the
mass loading of the outflow may be reduced, and the Alfv\'en radius may increase,
leading to a stronger centrifugal acceleration of the flow. As the faster parts of 
the fireball catch up with the earlier ejected slower part, a strong shock develops
at a radius $r\sim v\tout\sim 10^{13}v_9\tout_4$~cm. 

The development of shocks can be affected by the magnetic field carried by 
the fireball. The field is transverse to the outflow velocity at radii $r\gg R$,\footnote{
     Magnetic flux conservation in the expanding flow gives the transverse field 
     $B_\perp\propto r^{-1}$ and the radial field $B_r\propto r^{-2}$.       }
and radial waves propagate in the plasma with the fast magnetosonic speed 
$v_m\approx \vA\simlt v$, where $\vA=B(4\pi \rho)^{-1/2}$.
Shock dissipation will be suppressed if variations in the ejection velocity are
smaller than $v_m$, which can be satisfied when a large fraction of the
fireball energy is carried by the magnetic field. The field itself can, however,
dissipate through magnetic reconnection in the outflow (Drenkhahn \& Spruit 2002),
providing an alternative source of heat.

Internal dissipation at radii $r\gg R$ significantly increases the expected luminosity.
The adiabatic cooling factor at the diffusion radius $(\Rd/R)^{-2/3}\sim 10^{-3}$
is offset by heating and should be replaced by a larger factor $\sim 10^{-2}-10^{-1}$.
The dissipative fireball can easily emit 
$L\sim 10^{42}$~erg~s$^{-1}$ at $\Rd$ and possibly a comparable luminosity 
at $\Rph$.

The luminosity generated at the photospheric radius $\Rph$ may also be 
estimated as
\beq
   \Lph\sim \frac{\effph\Eout}{\tph}, 
\eeq
where $\effph$ is the dissipation efficiency at $r\sim\Rph$ and $\tph\sim \Rph/v$ is 
the characteristic timescale of photospheric emission. This gives,
\begin{eqnarray}
\nonumber
  \Lph &\sim& \effph\,\frac{\Mout v^2}{2}\,\frac{v}{\Rph}
  \sim \pi^{1/2}\effph\left(\frac{\Mout}{\kappa_\star}\right)^{1/2} v^3 \\
   &\approx& 6\times 10^{42}\, \effph\, \left(\frac{\kappa_\star}{\kappaT}\right)^{-1/2}
     \Mout_{-3}^{1/2} v_9^{3} {\rm ~erg~s}^{-1}.
\end{eqnarray}
A moderate dissipation efficiency $\effph$ of a few per cent provides a high 
photospheric luminosity $\Lph>10^{41}$~erg~s$^{-1}$.

\subsection{Detailed models with accurate opacity}

The above estimates were scaled to Thomson opacity 
$\kappaT\approx 0.2$~cm$^2$~g$^{-1}$. The actual opacity of the outflowing 
plasma is a function of temperature $T$ and density $\rho$.
This dependence can be found e.g. in the OPAL tables 
(Iglesias \& Rogers 1996) for helium or carbon-oxygen composition. 
Using the OPAL tables, we have calculated several simple models 
of the outburst with accurate opacities. The models assumed the 
following mass loss rate of the remnant,
\beq
  \dM=10^{26} {\rm ~g~s}^{-1} \times \left\{\begin{array}{ll}
     1 & t<\tout \\
      (t/\tout)^{-\beta} & t>\tout
                  \end{array}
          \right.
\eeq
At radii $r<\Rd$, radiation carried by the flow is described by $L=4\pi r^2 U v$, 
where $U=aT^4$. It is approximated by 
\beq
     L\approx \frac{\dM v^2}{2} \, r_9^{-\delta}, \qquad r<\Rd.
\eeq
Outside the diffusion radius $\Rd$, radiation density $U$ is quickly reduced
so that luminosity $L\approx 4\pi r^2 U v_{\rm diff} \approx const$, where 
$v_{\rm diff}$ increases from $v$ at $\Rd$ to $\sim c$ at the photoshere $\Rph$.

We calculated the models with outflow velocity $v=10^{9}$~cm~s$^{-1}$,
$\tout=10^4$-$10^5$~s, and studied two cases for $\delta$: 
$\delta=2/3$ (simple adiabatic cooling, no internal dissipation) and 
$\delta=1/3$ (toy model where cooling is slowed down due to internal dissipation).
Accurate opacities $\kappa(\rho,T)$ were used in the calculations, for two 
chemical compositions: pure helium or carbon-oxygen 
(with the carbon mass fraction $X_C=0.4$). 
The results confirmed that 
$\kappa$ is close to $\kappaT$ (within $\sim 50$ per cent) until the 
fireball reaches $r\sim (2-6)\times 10^{13}$~cm
where its temperature decreases to $\sim 10^4$~K; then the opacity drops.
The diffusion radius $\Rd\simlt 10^{14}$~cm and photospheric radius 
$\Rph\simgt 10^{14}$~cm are close to the estimates in Section~3.1.
Note that all absorptive processes --- free-free,
bound-free, and bound-bound --- contribute to the fireball opacity
in the main emission region $10^{13}\simlt r \simlt 10^{14}$~cm.
This should help the thermalization of escaping radiation.

\subsection{Decay of the outburst}

At times $t>\tout$ the mass loss rate of the merger remnant $\dM$ decreases 
and the outburst decays. At late times, the diffusion radius $\Rd$ and the 
photospheric radius $\Rph$ are both reduced proportionally to $\dM$.
The bolometric luminosity scales as $L\propto \dM v^2 \eta$.  This gives 
the effective temperature of escaping radiation 
$T\propto \dM^{-1/4} v^{1/2} \eta^{1/4}$, 
i.e. the temperature grows if the efficiency $\eta$ and the flow velocity $v$ 
remain approximately constant. 

Then \Eq~(\ref{eq:LO}) gives the optical luminosity,
\beq
  L_O\propto \dM^{7/8} v^{1/2} \eta^{1/4}.
\eeq 
The optical luminosity is sensitive to the mass loss rate of the remnant and 
should drop when $\dM$ is reduced.

\subsection{Nonthermal emission}

Internal shocks dissipate the velocity variations $\Delta v\simlt v$ and heat 
the plasma to temperature $kT\sim 10^2$~keV. The plasma
immediately converts its heat to radiation behind the shock (via inverse
Compton scattering, bremsstrahlung, and line emission). 
Thus, an X-ray luminosity $L_X$ up to $\sim 10^{42}$~erg~s$^{-1}$
is generated inside the fireball. As long as the fireball is opaque to X-rays, 
$L_X$ is re-processed into quasi-blackbody radiation. Late internal shocks 
propagating outside the X-ray photosphere could produce observable 
X-ray emission. Internal shocks may also generate non-thermal particles, which 
produce high-energy (inverse Compton) photons and synchrotron radiation
with a broad spectrum. 

At early times the remnant age is smaller than the radiation escape time, 
$t<\Rd/v$, so internal shocks only occur in the highly opaque zone. 
The longest timescale of the central engine variability $t_{\rm var}$ 
is comparable to its age, and the growing age of the remnant helps formation 
of internal shocks at large radii. 
At the same time, the photospheric radius of the outflow is reduced 
$\propto \dM$ as the mass loss rate decreases. 
Thus, propagation of shocks in the transparent zone 
becomes more likely at late times, after the peak of the outburst.
Then highly variable nonthermal emission may be detected in addition
to the quasi-thermal component.

%#################################################################

 \section{Outflow from a quiet corona}

This section briefly discusses the mass loss rate that could be expected from 
a ``quiet'' corona, i.e. in the absence of flares due to field-line twisting by 
footpoint motion. 
The corona may become relatively quiet as the merger remnant
ages past $\tvisc$. Shen et al. (2012) argue that the remnant should 
evolve into a quasi-spherical object in approximately solid-body rotation
(although we do not exclude that the aging remnant is still surrounded by 
a low-mass Keplerian disc).
The viscously heated remnant expands, and its maximum (break-up)
angular velocity decreases as $\Omega_{\max}\propto R^{-3/2}$ while 
the actual angular velocity decreases as $\Omega\propto R^{-2}$ 
(as long as its angular momentum is approximately conserved).

Even in the absence of differential rotation and coronal flares, 
the rotating magnetized remnant would lose mass along open magnetic field lines. 
Some of the field lines must be open by rotation.
In the force-free approximation, the minimum open magnetic flux is 
$\Psi\sim\mu/\RLC$, where $\RLC=c/\Omega$
is the light cylinder radius and $\mu$ is the magnetic dipole moment of the 
remnant. Then less than one per cent of magnetic field lines are open. 
More field lines can become open if significant mass is lifted from the 
remnant and its corona is changed from the force-free configuration
(Mestel \& Spruit 1987).

\subsection{Mass flux}

Consider the quiet corona supported by radiation pressure in a fixed strong 
magnetic field. The field enforces co-rotation of 
the plasma with angular velocity $\Omega$. To demonstrate the reason of mass 
loss and estimate its magnitude consider a field-line bundle near the equatorial 
plane in the simplest, monopole-like geometry (radial field lines). The plasma
at the base of the corona must be close to hydrostatic equilibrium.
The hydrostatic balance for the radiation-dominated plasma reads
\beq
\label{eq:hb}
    \frac{\kappa\,F}{c}=\geff(r)=\frac{GM}{r^2}-\Omega^2\,r,
\eeq
where $F$ is the radiation flux along the magnetic field line and 
$\kappa\approx\kappaT={\rm const}$ is the plasma opacity.
The hydrostatic balance implies continual heating of the plasma,
\beq
\label{eq:dQ}
   \dot{Q}=-\frac{1}{r^2}\frac{\partial}{\partial r}
     \left(r^2\,F\right)=3\frac{c}{\kappa}\,\Omega^2.
\eeq
Hence, a steady state is possible only if the plasma is 
gradually flowing away from the object, which modifies the hydrostatic picture
and introduces advection of heat (dominated by radiation) by the opaque flow.

The hydrostatic approximation~(\ref{eq:hb}) is still useful where the flow velocity 
is subsonic. As the sound speed $c_s$ of the heated flow increases, 
it climbs the effective (gravitational $-$ centrifugal) potential barrier, and its 
hydrostatic scale-height grows,
\beq
\label{eq:H}
   H(r)=\frac{c_s^2(r)}{g(r)}.
\eeq
At sufficiently large altitudes, the heated flow passes through the sonic point 
and escapes with velocity $v\sim c_s\sim v_0$. However, to estimate the mass
loss rate one can consider the deep, subsonic, approximately hydrostatic region 
where $v\ll c_s\ll v_0$ and $H\ll R$, $r\approx R$.
The heating timescale $\sim U/\dot{Q}$ is comparable to the timescale for
the change in $H$, 
\beq
\label{eq:cond}
    \frac{3\rho c_s^2}{\dot{Q}}\sim \frac{H}{v}.
\eeq
\Eqs~(\ref{eq:dQ})-(\ref{eq:cond}) give the mass flux in the outflow,
\beq
  \Fm=\rho v\sim\frac{\dot{Q}}{3g}=\frac{c\Omega^2}{\kappa g}
     =\frac{c}{\kappa R}\,\left(\frac{\Omega_{\max}^2}{\Omega^2}-1\right)^{-1}.
\eeq
The net mass loss rate is $\dM\sim \Fm A$, where $A$ is the area
of the footprint of the open field-line bundle on the remnant. 
This mass loss rate is much smaller than that of the active, flaring corona 
discussed in Sections~2 and 3.

 \subsection{Strong field regime}

In a sufficiently strong 
magnetic field, the outflow remains magnetically dominated and continues 
to accelerate centrifugally at $r\gg R$, reaching the light cylinder with 
$v\approx c$.
In this regime, most of the energy lost by the rotating object is carried 
away by the Poynting flux and described by the standard pulsar spindown formula,
\beq
\label{eq:sd}
   \dEem\sim \frac{\mu^2\Omega^4}{c^3}
    \sim 4\times 10^{38} \,\mu_{37}^2\, \Omega_{-1}^4
   {\rm ~erg~s}^{-1},
\eeq
where $\mu\sim BR^3$ is the magnetic dipole moment of the remnant. 
The kinetic power of the matter ejected at the light cylinder is $\dEm\sim \dM c^2$;
more energy is transferred from the Poynting flux to the plasma outside $\RLC$ 
where $\dEm$ grows to its asymptotic value (Michel 1969),
\beq
  \dEm\sim \sigma^{1/3} \dM c^2, \qquad \sigma=\frac{\dEem}{\dM c^2}>1.
\eeq
Estimating $A/4\pi R^2\sim R/\RLC$ (which would roughly correspond to a dipole
field), one obtains
\begin{eqnarray}
\nonumber
   \dEm &\sim& \LE \,\sigma^{1/3} \frac{R}{\RLC} \frac{c^2}{v_0^2}
   \sim  \LE\, \sigma^{1/3} \frac{R^2}{r_g\RLC} \\
   &\sim& 3\times 10^{39}\,\sigma^{1/3}\,R_9^2\,\Omega_{-1}
     {\rm ~erg~s}^{-1},
\label{eq:mat}
\end{eqnarray}
where $r_g=GM/c^2\sim 10^5$~cm and 
$\LE=4\pi GM c/\kappa\approx 2.5\times 10^{38}(M/M_\odot)$~erg~s$^{-1}$
is the Eddington luminosity with $\kappa=\kappaT$. Comparing 
\Eqs~(\ref{eq:sd}) and (\ref{eq:mat}), one can see that the condition $\sigma>1$ 
(the strong-field regime) is satisfied when $\mu_{37}^2\Omega_{-1}^3>10$.

Radiation is trapped and advected by the outflow below the sonic radius 
$r_s\sim R$. 
At larger radii, radiation diffusion becomes faster than advection; here radiation 
diffuses through the outflow and escapes at its photosphere, which is located 
well inside of the light cylinder (roughly at $r\sim 10^{10}$~cm). A moderate 
quasi-thermal luminosity is emitted from the object, comparable to the 
Eddington luminosity $\LE\sim 10^{38}$~erg~s$^{-1}$, with the effective 
temperature comparable to $10^5$~K. 
The kinetic and magnetic power of the outflow in the strong-field regime significantly
exceeds the quasi-thermal photospheric luminosity. It may be partially dissipated 
and converted to radiation at large radii, producing nonthermal radiation with a 
broad spectrum.

\subsection{Weak field regime}

The regime $\dEem>\dEm$ discussed in Section~4.2 is satisfied if the surface 
magnetic field $B\sim \mu R^{-3}$ exceeds the characteristic value
\beq
    B_1\sim \left(\frac{2\pi \Fm c^4}{R^3\Omega^3}\right)^{1/2}
    \sim \left(\frac{2\pi c^5}{\kappaT R^4\Omega^3}\right)^{1/2}
    \sim \frac{3\times 10^{10}}{R_9^{2}\,\Omega_{-1}^{3/2}} {\rm ~G.}
\eeq
If the surface field is weaker, the Alfv\'en radius $\RA$ 
(at which $B^2\sim 4\pi \rho v^2$) becomes smaller than $\RLC$, 
the open field-line bundle becomes broader (its footprint area $A$ is increased)
and the asymptotic outflow velocity is reduced below $c$. 

The calculation of $\Fm$ in \Sect~4.1 is valid if the surface magnetic field is 
strong enough to enforce co-rotation of the coronal plasma.  This condition 
requires a minimum field,
\beq
   B_0\sim \left(4\pi \Fm v_0\right)^{1/2}\sim 10^6 \frac{\omega}{(1-\omega^2)^{1/2}} 
       R_9^{-3/2}\, {\rm ~G},
\eeq
where $\omega=\Omega/\Omega_{\rm max}$.
If $B\sim B_0$ then $\RA\sim R$, $\vA\sim v_0$, and a large fraction of 
magnetic field lines are open by the outflow, $A\sim R^2$. Then 
$\dot{M}\sim R^2\Fm\sim\ cR/\kappaT$ and $\dEm\sim \dM v_0^2/2$.

The scaling of $\RA$, $\vA$, $A$, $\dM$ and $\dEm$ with $B$ in the range of 
$B_0<B<B_1$ may be estimated by comparing their values at $B\sim B_0$
and $B\sim B_1$.\footnote{Results of more accurate calculations would 
     depend on details of 
     the magnetic configuration, in particular, the angle between the magnetic 
     dipole moment and the rotation axis of the object, and the presence of
     multipoles.}
This gives a rough estimate,
\beq
    \dM\propto B^{-1/2}, \qquad \dEm\propto B^{1/2}, \qquad (B_0<B<B_1).
\eeq
The maximum mass loss of the quiet corona, 
$\dM\sim cR/\kappaT\sim 10^{20}$~g~s$^{-1}$ 
is approached when $B\sim B_0$, with a modest energy output 
$\dEm\sim\dEem\sim 10^{38}$~erg~s$^{-1}$.

%#################################################################

\section{Discussion}

The active corona of the differentially rotating remnant produces an outflow with the 
velocity $v\sim 10^9$~cm~s$^{-1}$.
The expected duration of the high mass-loss phase is  
$\tout\simgt 10^4$~s, comparable to the lifetime of strong differential 
rotation in the object. 
The outflow caries away a fraction of the energy stored in differential rotation.
This fraction is uncertain. Numerical simulations of Ji et al. (2013) and simple 
estimates suggest that it can be comparable to one per cent, which corresponds 
to the ejected mass $\Mout\sim 10^{-3}M_\odot$.
In this paper we discussed the consequences of this mass ejection. 

The outflow creates a dense fireball of characteristic thickness 
$v\tout\sim 10^{13}$~cm,
which becomes transparent to radiation at the 
photospheric radius $\Rph\sim 10^{14}$~cm. It is unlikely to be spherically 
symmetric, and may contain a faster jet near the rotation axis. 
The initial thermal energy of the fireball is dominated by radiation 
and comparable to its kinetic energy $\Eout\sim \Mout v^2/2$; 
most of it is lost to adiabatic cooling.
In the absence of dissipative processes in the fireball, only a small fraction 
$\sim 10^{-3}$ of $\Eout$ is radiated away. Even this small fraction 
would give an interesting transient event with luminosity 
$L\sim 10^{40}-10^{41}$~erg~s$^{-1}$.

We further argued that the outflow from the active corona 
must be highly variable in a broad range of timescales.
The variable  outflow is expected to develop internal shocks 
(in addition to possible delayed magnetic dissipation), which convert
a fraction of the fireball kinetic energy to heat, offsetting adiabatic cooling. 
The dissipative fireball is expected to produce a higher luminosity 
$L\sim 10^{41}-10^{42}$~erg~s$^{-1}$, comparable to the luminosities of 
core-collapse supernovae. 
The peak timescale of the produced outburst is between $\tvisc$ and 
$\Rph/v$, comparable to 1~day.

The emitted radiation has the effective temperature 
$T\simgt 10^4$~K, and a significant fraction of the outburst is 
emitted in the optical band. Estimates in this paper used the simplifying 
blackbody assumption up to the photospheric radius. 
A more realistic emission spectrum could be 
obtained with detailed transfer calculations.

The fireball emission is $\sim 10^3$ times brighter than classical novae, and 
may be called ``kilonova'', similar to the transients expected from neutron-star 
mergers (e.g. Metzger et al. 2010; Kasen et al. 2013). The classical novae 
are emitted by ejecta of a smaller mass $\Mout\sim 10^{-5}-10^{-4}M_\odot$ 
moving with velocity $v\sim 10^8$~cm~s$^{-1}$.
Their photospheric radii approach $\Rph\sim 10^{13}$~cm when
the ejecta temperature decreases to $\sim 10^4$~K; then the ejecta opacity 
quickly decreases, and the photosphere recedes (e.g. Gallagher \& Starrfield 1978). 
Similar behavior, but with a larger $\Rph\sim 10^{14}$~cm and higher $L$, 
may be expected for the outbursts from the WD mergers.
We emphasize that the proposed outburst mechanism does not 
invoke nuclear reactions, in contrast to classical novae, 
supernovae or kilonovae from NS mergers. 
The main energy source is differential rotation that is partially converted to heat 
around the remnant through magnetic dissipation.

A special feature of this scenario is that the decay of the central engine 
activity occurs on a timescale comparable to the radiation diffusion time
$\Rd/v\sim 3\times 10^4$~s.   
This coincidence implies that the observed light curve can be affected by both 
the fireball expansion and the central engine evolution. The long timescale
variability of the central engine also makes it possible for internal
dissipation to occur in an extended range of radii not much below the main
emission zone $r\sim 10^{13}-10^{14}$~cm. A similar mechanism is unlikely to 
work for the mildly relativistic outflows from NS mergers, which are ejected
on a timescale $<10^2$~s, much shorter than the 
emission time $t\sim 10^5$~s.\footnote{Internal shocks and magnetic 
      dissipation can be efficient in the ultra-relativistic jets producing 
      GRB emission, because in such jets the dissipation radius can be 
      comparable to or exceed the photospheric radius (e.g. Piran 2004).}

The outbursts with luminosities $L\sim 10^{41}-10^{42}$~erg~s$^{-1}$
(absolute visual magnitudes of $-13$ to $-16$) can easily be detected 
in optical surveys with sufficiently short cadence $\sim 1$~day. 
Adopting that the observed volume 
(and the expected number of detections) scales as $L^{3/2}$ and the 
rate of WD mergers is comparable to that of thermonuclear supernovae
(Badenes \& Maoz 2012),
one can roughly estimate the expected number of detections as 
$N\sim (L/L_{\rm SN})^{3/2} N_{\rm SN}$, where $N_{\rm SN}$ is the 
number of detected thermonuclear supernovae and 
$L_{\rm SN}\sim 10^{43}$~erg~s$^{-1}$ is their luminosity.
The current survey by the Palomar
Transient Factory may detect the outbursts, and the upcoming 
Large Synoptic Survey Telescope should routinely observe them.

Magnetically powered outbursts from WD mergers should differ from normal 
supernovae in several respects.
(i)~The optical light curve should peak early (one day timescale) and then 
show an unusual decay that reflects the decay of differential rotation in the remnant.
(ii) The effective temperature of emission decreases as the optical luminosity
rises towards its peak, reaches the minimum $T_{\min}\approx 10^4$~K near 
the peak, and is expected to grow while the source is fading. 
(iii) Line features in the spectrum should differ from those in supernovae, 
as the fireball is dominated by unburned material, with chemical composition 
close to that of the merging WDs (dominated by carbon, oxygen or helium). 
(iv) Heating by internal shocks and delayed magnetic dissipation might extend 
to large radii where the fireball becomes transparent. Then a variable nonthermal 
component may be emitted with a broad spectrum, from radio to gamma-rays 
(Section~3.6).

Several transients with luminosities of $10^{41}$-$10^{42}$~erg~s$^{-1}$,
fast decay and puzzling chemical composition have recently been detected; 
they were interpreted as unusual, rare variations of supernovae, although their 
origin is not established (see e.g. Kleiser \& Kasen 2013 and refs. therein). 
Magnetically powered fireballs from WD mergers should produce similar 
events, contributing to the diversity of observed transients.

The decay of differential rotation must significantly 
change the observational appearance of the merger remnant at 
late times $t\gg\tvisc$. 
The corona becomes less active and the mass loss should significantly decrease
(Section~4). The strong magnetic fields generated by differential rotation,
$B\sim 10^{10}-10^{11}$~G, are expected to decay.
The surviving fields may still be relatively strong, $B\sim 10^8-10^9$~G, 
providing a possible formation scenario for magnetic white dwarfs 
(Garc\'ia-Berro et al. 2012; K\"ulebi et al. 2013). 
The remnant can temporarily increase in size, due to viscous 
heating, and correspondingly slow down its rotation (Shen et al. 2012); 
then the remnant will cool down on the long Kelvin-Helmholtz timescale and shrink.

The ejected fireball will eventually be decelerated by the surrounding medium. 
The characteristic deceleration radius $\Rdec$ is where the swept-up external 
mass becomes comparable to the ejecta mass $\Mout$,
\beq
  \Rdec\sim 7\times 10^{17} \Mout_{-3}^{1/3} n^{-1/3} {\rm ~cm},
\eeq 
where $n$ is the number density of the external medium in units of cm$^{-3}$.
The ejected mass $\Mout$ reaches $\Rdec$ in time 
$t_{\rm dec}\sim\Rdec/v\sim 20\,v_9^{-1}\Mout_{-3}^{1/3}\,n^{-1/3}$~yr.
This interaction is accompanied by a strong shock wave, which 
is expected to produce nonthermal particles and synchrotron radiation. 
Significant radio emission may be produced at this stage.

\section*{Acknowledgements}

I thank Anthony Piro for discussions of white dwarf mergers in the spring of 
2012, and the Astronomy Department of California Institute of Technology for 
hospitality during my visit supported by the Merle Kingsley fellowship.
I thank Romain Hasco\"et, Brian Metzger, Henk Spruit, Indrek Vurm and 
the referee for comments on the manuscript.

%#############################################################
% \bibliographystyle{apj}
% \bibliography{apj-jour,merger}

% \bibliographystyle{mn2e}
% \bibliography{bib.bbl}

\end{document}